\documentclass[apjl]{emulateapj}
\usepackage{hyperref}
\hypersetup{
colorlinks=true,
linkcolor=red,
citecolor=blue,
filecolor=cyan,
urlcolor=magenta,
}
\usepackage{graphicx}
\usepackage{natbib}
\usepackage{multirow}
\usepackage{subfigure}

\begin{document}
\title[Detecting gravity modes in the solar $^8B$ neutrino flux]{
Detecting gravity modes in the solar $^8B$ neutrino flux}
\author{Il\'idio Lopes\altaffilmark{1,2} and Sylvaine Turck-Chi\`eze\altaffilmark{3}}
\altaffiltext{1}{Centro Multidisciplinar de Astrof\'{\i}sica, Instituto Superior T\'ecnico, 
Universidade de Lisboa, Av. Rovisco Pais, 1049-001 Lisboa, Portugalilidio.lopes@ist.utl.pt;ilopes@uevora.pt} 
\altaffiltext{2}{Departamento de F\'\i sica, Escola de Ciencia e Tecnologia, 
Universidade de \'Evora, Col\'egio Luis Ant\'onio Verney, 7002-554 \'Evora, Portugal} 
\altaffiltext{3}{CEA/IRFU/Service d'Astrophysique, CE Saclay, F-91191 Gif sur Yvette, France; sylvaine.turck-chieze@cea.fr}  
\date{\today}
\begin{abstract} 
The detection of gravity modes produced in the solar radiative zone has been a challenge in modern astrophysics for more than 30 yr and their amplitude in the core is not yet determined.   
In this Letter, we develop a new strategy to look for standing gravity modes through solar neutrino fluxes. We note that due  to a resonance effect, the gravity modes of low degree and low order have the largest impact on the $^{8}B$ neutrino flux. The strongest effect is expected  to occur for the dipole mode with radial order $2$, corresponding to periods of about 1.5 hr. These standing gravity waves produce temperature fluctuations that are amplified by a factor of 170 in the boron neutrino flux for the corresponding period,
in consonance with the gravity modes. 
From current neutrino observations, we determine that the  maximum temperature 
variation due to the gravity modes in the Sun's core is smaller than  $5.8\times 10^{-4}$.
This study clearly shows that due to their high sensitivity to the temperature, the $^8B$ neutrino flux time series  is an excellent tool to determine  the properties of gravity modes in the solar core. Moreover, if gravity mode footprints are discovered in the $^{8}B$ neutrino flux,  
this opens a new line of research to probe the physics of the solar core as non-standing gravity waves of higher periods cannot be directly detected by helioseismology but could leave their signature on boron neutrino or on other neutrino fluxes.
\end{abstract}


\maketitle

%
%
\section{Introduction}

Neutrinos are the most direct evidence of  nuclear reactions occurring in the core of the Sun.
For two decades now, neutrinos have complemented helioseismology 
in diagnosing  the standard and non-standard structure of the Sun ~\citep[e.g.,][]{1993ApJ...408..347T,2011RPPh...74h6901T,2012RAA....12.1107T}. The last decade have been particularly rich as the statistics obtained with the neutrino detectors  has impressively increased together with the knowledge of their oscillation behavior between different species. These facts push us to study 
whether neutrinos could investigate the dynamics of the solar core even better than helioseismology, even though using neutrinos as solar core probes presents a number of challenges in itself. This is due to the  complex physical processes occurring inside the 
Sun, but also due their still unknown basic properties~\citep[e.g.,][]{2008PhR...460....1G}.  

In a star like the Sun, it is expected that gravity modes (or $g$-modes)
 should affect the evolution of the star by influencing
the transport of angular momentum 
and the transport of energy in the radiative region ~\citep{1997ApJ...475L.143K,1997A&A...322..320Z,2002ApJ...574L.175T,Mathis2008}.
Although their existence is almost certain from a theoretical point of view, 
their detection is a long-awaited goal of helioseismology~\citep[e.g.,][]{2004ApJ...604..455T,Garcia2007}.
These modes provide a better diagnostic of the Sun's core for the sound speed and  deep rotation than the more easier  observed standing acoustic waves (p-modes), but remain difficult to detect at the surface where only few of them seem to have imprinted the space GOLF/$SOHO$ instrument until  now.

The high sensitivity of $g$-modes to the nuclear region is a consequence  
of their eigenfunctions having large amplitude in the core and low amplitude at the surface,
which is the opposite  of acoustic modes~\citep{1989nos..book.....U}.
The evanescent nature of $g$-modes in the convection zone adds to the complexity of the transition layer between radiation and convection and makes them difficult to observe at the solar surface~\citep{1990A&A...227..563B}.
The current models suggest that $g$-modes have very low velocity amplitudes 
in the photospheric layers, smaller than 
$5\;  {\rm\; mm \; s^{-1}}$~\citep{1996ApJ...458L..83K,2009A&A...494..191B},  
which is much smaller than the maximal acoustic mode velocity amplitudes of  
 $20  {\rm\; cm\;  s^{-1}} $~\citep[e.g.,][]{1995A&A...293...87K}.
So, the former modes are much more difficult to observe than the latter ones.

If $g$-modes provide a crucial tool to study the physical processes present in the solar core, 
they could also be used as a cosmological probe to validate the existence of dark matter inside the Sun
~\citep{2012ApJ...752..129L,2010Sci...330..462L,2010ApJ...722L..95L,2010PhRvD..82h3509T,2010PhRvD..82j3503C,Turck2012},
or in Sun-like stars~\citep{2013ApJ...765L..21C} and neutron stars~\citep{2012PhRvL.108s1301K,2011PhRvD..83h3512K,2011PhRvL.107i1301K,2010PhRvD..82f3531K}.

This study is of particular interest because several solar neutrino detectors~\citep{2011PhRvD..83e2010A,2011arXiv1107.2901A,2010PhRvD..82c3006B}
have been  measuring  $^8B$  neutrino fluxes for a long time and new ones are scheduled.
\citet{2013ApJ...777L...7L} has already computed the impact of global acoustic and f-modes 
on the $^8B$  neutrino flux but their effect in the core should be small as their amplitude is maximal at the surface except for the low-frequency part, which has a mixed character.
In this Letter, we predict for the first time how 
the different standing gravity waves influence the $^8B$ neutrino flux because these modes have the maximum of their amplitude near the center. Furthermore,
we show which  gravity oscillation modes will be easier to detect.   

\begin{figure}
\centering
\includegraphics[scale=0.30]{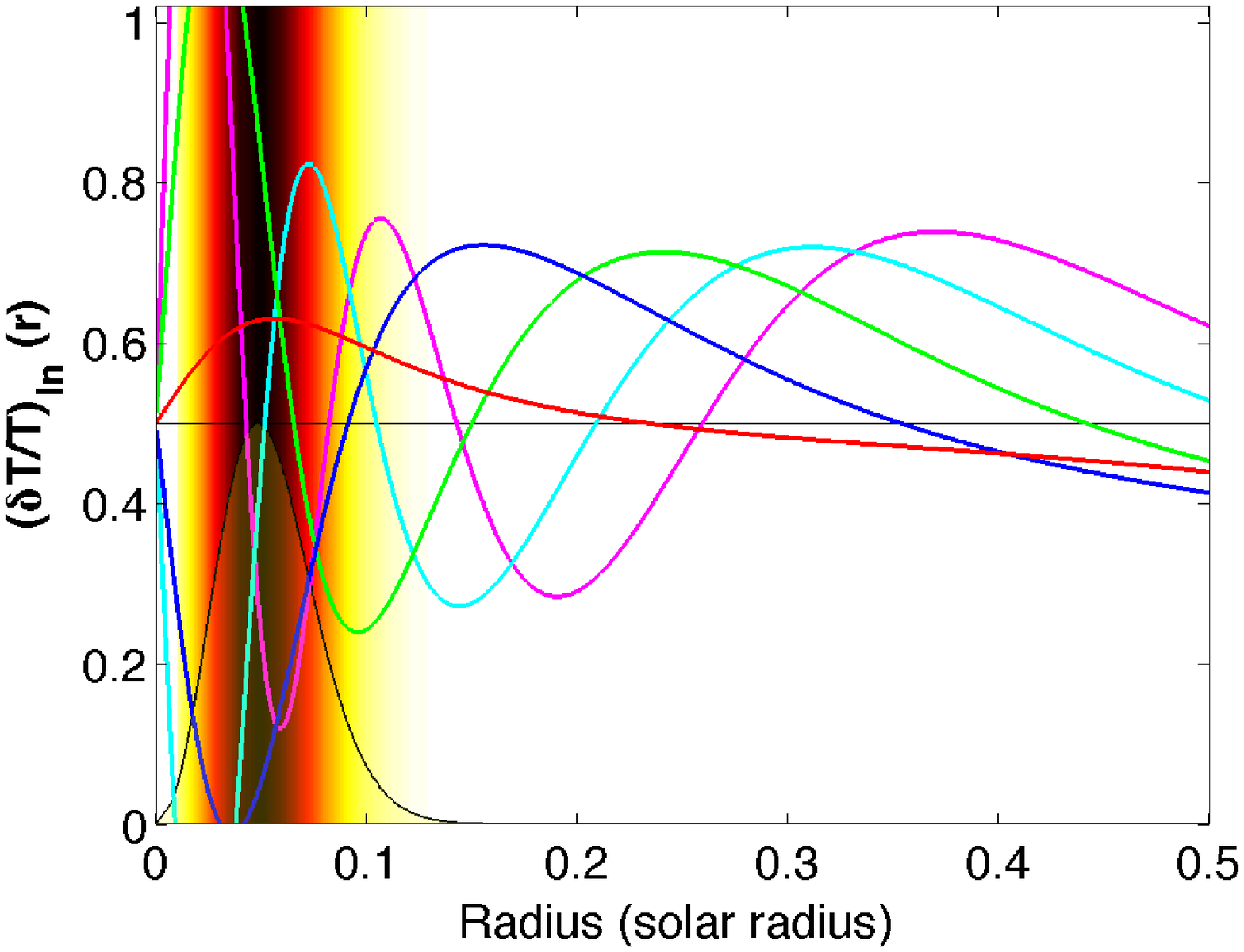}
\includegraphics[scale=0.30]{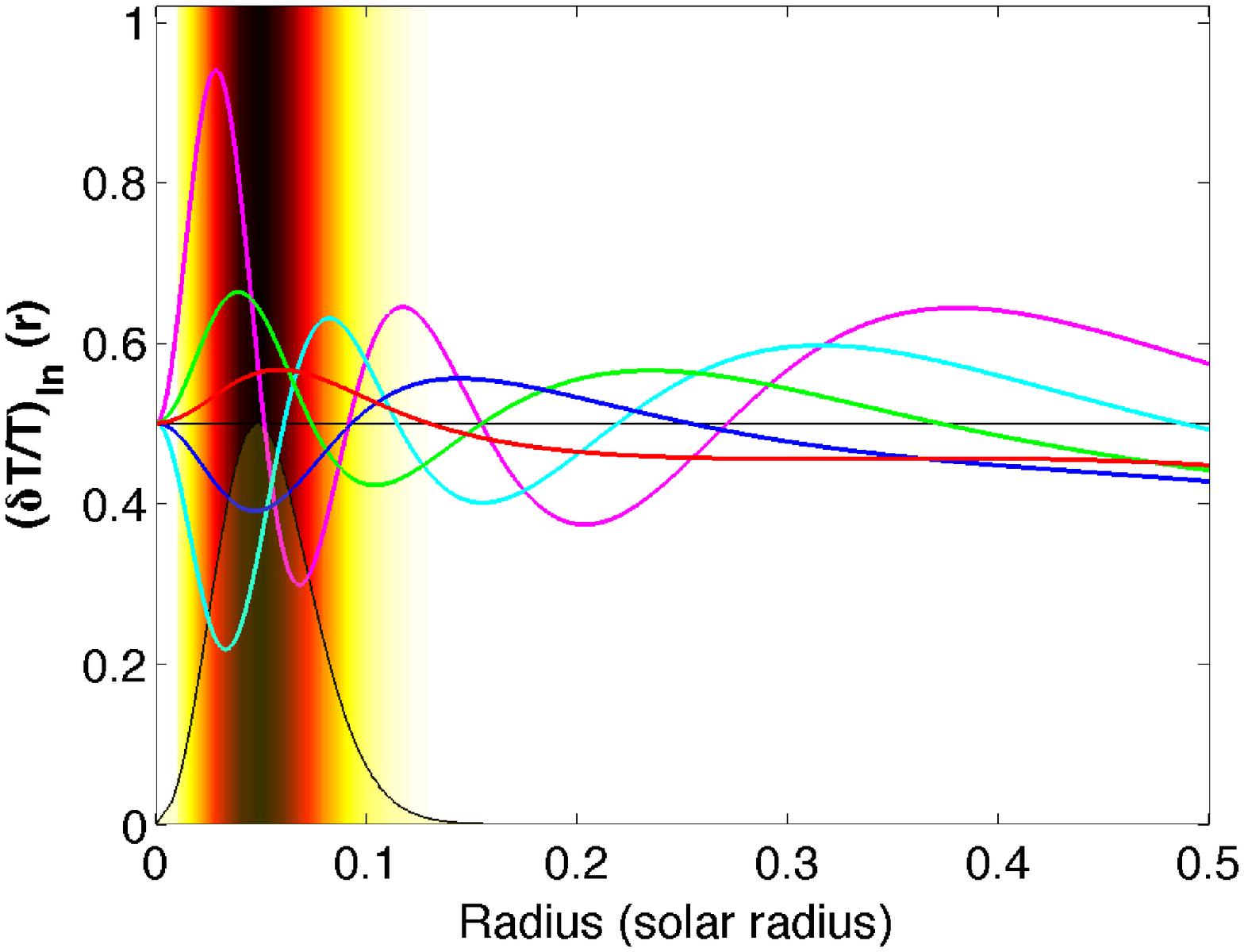}
\caption{ 
$\left(\delta T/T\right)_{\l n}$ eigenfunctions for dipole $\l=1$ (top) and quadrupole (bottom) $g$-modes: 
$n=-1$ (red curve), $n=-2$ (blue curve), $n=-3$ (green curve), 
$n=-4$ ( cyan curve) and $n=-5$ (magenta curve) deduced from the solar standard model.
For convenience, all eigenfunctions
are normalized to 0.5 which corresponds to twice the minimum of $\left(\delta T/T\right)_{1,-2}$.
The radial $^8B$ neutrino flux $\phi (r)$ is superimposed (yellow area and black contour). 
}
\label{fig:1}
\end{figure}

\section{Perturbation of the total neutrino flux}

Any oscillation is regarded as a perturbation of the equilibrium which is expressed as a superposition of propagating waves, i.e. a combination of standing waves or eigenmodes of vibration. Any thermodynamic quantity, like the temperature $T$, is perturbed relative to its
equilibrium value $T_o$, such that  $T(r)=T_o(r) +(\delta T)_{\l n} (r) $, where  $ (\delta T)_{\l n} (r) $ denotes the temperature eigenfunction corresponding to the mode of degree $\l$ and order $n$. 
One denotes $\omega_{\l n} $ and $\eta_{\l n}$ the related  eigenfrequency and damping rate~\citep{1989nos..book.....U}.
The eigenfrequencies and eigenfunctions are computed numerically 
for the solar standard model~\citep[SSM in Table~\ref{tab:1} from][]{2013ApJ...765...14L}.
This model is similar to others published in the
literature~\citep[e.g.,][]{2010ApJ...715.1539T,2010ApJ...713.1108G, 2011ApJ...743...24S},
which use the nuclear reaction rates from the NACRE compilation~\citep{1999NuPhA.656....3A}
and the solar mixture of~\citet{2009ARA&A..47..481A}. The structure model of the present Sun was computed using the stellar evolution code CESAM~\citep{1997A&AS..124..597M}
and the pulsation eigenmodes using the code adipls~\citep{2008Ap&SS.316..113C}.
Table~\ref{tab:1} shows the frequencies (freq.) of  lower $|n|$ $g$-modes, including the values of the seismic model specifically calculated with a detailed description of the core and with a sound speed in agreement with the seismic observations~\citep[called SeSM in Table~\ref{tab:1} from ][]{2007ApJ...668..594M}.
An accurate calculation of $g$-mode frequencies and eigenfunctions require a precise 
determination of the central boundary conditions. 
In particular, we have reduced the size of the shells (increased the number of shells) in the central region to describe properly the first 5\% of the solar radius.
We have verified that the equations 
are properly  solved near the center (particularly, near the singularities) 
and checked that the derivatives are computed correctly.  
We have also deduced the SeSM from the seismic observations, a sound speed profile in agreement with  the one derived from observations  to extract more precise neutrino fluxes and $g$-mode frequencies.
The SeSM sound speed changes are mainly due to opacity.
We follow the usual convention, in which $n$ is negative in the case of the $g$-modes.  
Figure~\ref{fig:1} shows some dipole $g$-mode eigenfunctions. 

The  propagation of gravity waves in the solar interior perturbs the local
thermodynamic structure, triggering fluctuations in the energy generation rate $\epsilon$
and on the solar neutrino fluxes.  
The energy generation rate per mass unit and time~\citep{1990sse..book.....K}
 between two reactants is such that $\epsilon \propto \rho\; \langle \sigma v \rangle $,
where  $\rho$ is the density and   $\langle \sigma v \rangle $ is the average cross section.
$\langle \sigma v \rangle $   contains the dependence of the temperature, $T$, such that
 $\langle \sigma v \rangle \sim T^{\mu}$, where  $\mu$ is the exponent.
The nuclear reaction rate responsible for the production of $^8B$ neutrinos,
$^7Be(p,\gamma)^8(e^{+}\nu)^8B^{*}(\alpha)^4He$ \citep[e.g.,][]{2011RPPh...74h6901T},
is very sensitive to temperature. Therefore, any local perturbation 
in the plasma will change the production rate of the neutrino flux. 
By comparing the $^8B$ neutrino fluxes of more than 1000 SSM computed inside the known uncertainties,
\cite{1996PhRvD..53.4202B} find that the $^8B$ neutrino flux $\phi$
is proportional  to $T_c^{24}$, where $T_c$ in the temperature at the center of the Sun.
This point has been also discussed in~\cite{1993ApJ...408..347T}.
Accordingly, we consider $\mu=24$ and write that
any oscillation mode will perturb the energy generation rate $\delta \epsilon$ 
by locally changing  the thermodynamic quantities:
\begin{equation}
\frac{\delta \epsilon}{\epsilon} = 
\frac{\delta \rho}{\rho} +\mu \frac{\delta T}{T}.   
\label{eq:a}
\end{equation}   

The perturbation of mass fraction of the reacting particles is negligible
so we have not taken them into account. 
Even though $\epsilon$  is generated by ppI and ppII chains, we examine here the
role of boron flux which corresponds to ppIII.

The total neutrino flux $\phi_{\l n}$, when perturbed by the eigenmode of
vibration with frequency $\omega_{\l n}$,  is  expressed as 
\begin{equation}
 \phi_{\l n}=\phi+ \Delta \phi_{\l n} (t)
\label{eq:phi_neutrino} 
 \end{equation} 
where $ \phi$ is the classic neutrino flux produced by 
the SSM (equilibrium model)
and $\Delta \phi_{\l n}(t)$ is the amount of neutrino flux varying with the eigenmode of vibration.
As the neutrino flux $\phi$ produced in this nuclear reaction is proportional to the generated energy $\epsilon$,
it follows that $\delta \phi / \phi  = \delta \epsilon/ \epsilon$. By integrating
Equation (\ref{eq:a}) for the total mass of the star,  $\Delta \phi_{\l n} (t)$ reads
\begin{equation} 
\Delta \phi_{\l n} (t)={\cal F}_{\l n}\phi \;e^{-i\omega_{\l n} t} \; e^{-\eta_{\l n} t} 
\label{eq:delta_phi_t}
\end{equation}  
where ${\cal F}_{\l n} = {\cal A}_{\l n}\;{\cal B}_{\l n} $, ${\cal A}_{\l n}$ is  
an amplitude related to the excitation source of gravity waves
and ${\cal B}_{\l n} $ is the fraction of the total neutrino flux 
perturbed by the eigenmode. ${\cal B}_{\l n} $ reads
\begin{equation} 
{\cal B}_{\l n}=\int_0^R \Psi_{\l n}(r)  dr 
\label{eq:B_ln}
\end{equation}
where $\Psi_{\l n}$ is the {\it gravito-neutrino eigenfunction}. It reads
\begin{equation} 
\Psi_{\l n}=C_\phi^{-1}\; 
\left( \frac{\delta T}{T}\right)_{\l n} \zeta \; \phi \; \rho\; 4\pi r^2 
\label{eq:Psi_ln}
\end{equation}
where $ C_\phi= \int \phi  (r) 4 \pi \rho (r) r^ 2  \;dr $ is a normalization constant, 
and $\zeta(r) = \left(\Gamma_3-1\right)^{-1} +\mu  $, where
$\Gamma_3-1=\left(\partial T /\partial \rho  \right)_s $ is the derivative being taken at constant specific entropy s. 
As the plasma could be considered as being completely ionized, $\Gamma_3=5/3$,
it follows that $\zeta(r)\equiv\zeta_o=3/2+\mu$.  If  $\mu=24$, accordingly $\zeta_o=25.5$. 
This result  shows that $\Psi_{\l n}$
is very sensitive to the temperature, although  not all eigenmodes are equally affected.  
\begin{figure}
\centering
\includegraphics[scale=0.35]{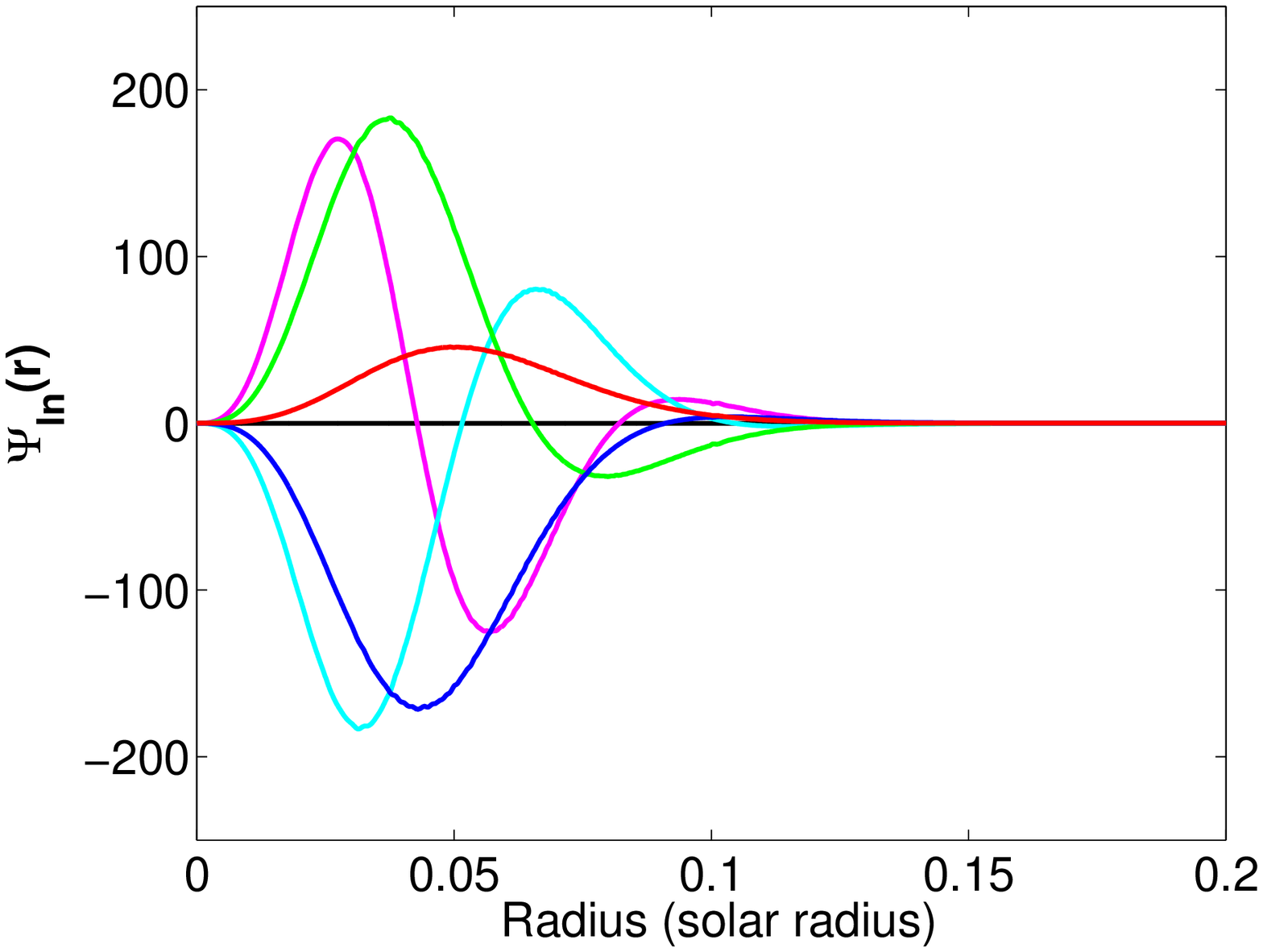}
\includegraphics[scale=0.35]{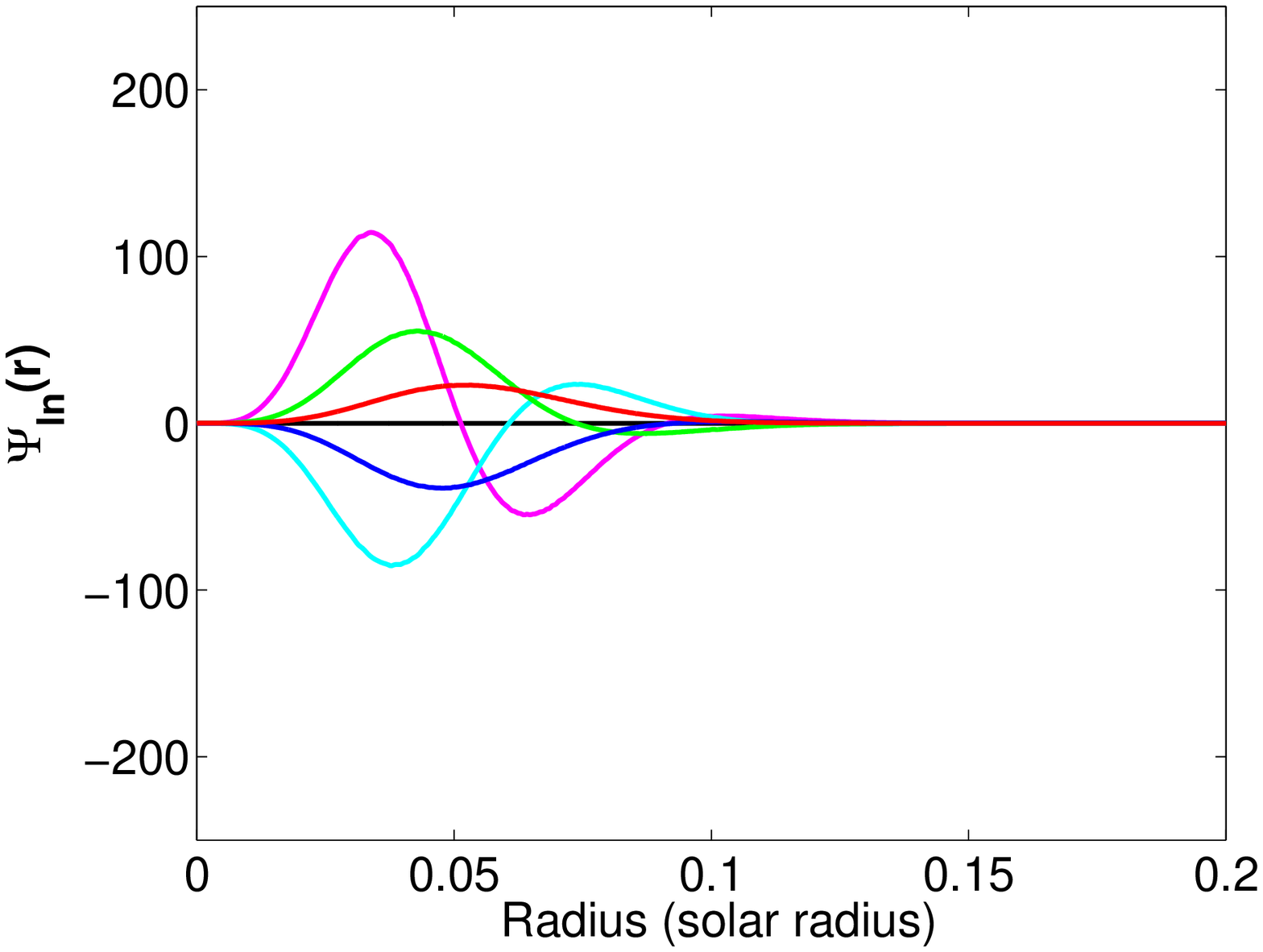}
\caption{
$\Psi_{\l n}$  are {\it gravito-neutrino  eigenfunctions} computed
for the selected  $g$-modes of Figure~\ref{fig:1} with $\zeta_o=1$ for dipole (top) and quadrupole (bottom) g- modes.
$\Psi_{\l n}$ is chosen dimensionless. }
\label{fig:2}
\end{figure}
  
\begin{figure}
\centering
\includegraphics[scale=0.3]{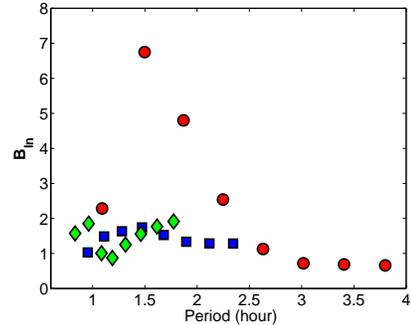}
\caption{$B_{\l n} $ computed for $g$-modes of degree $\l=1,2,3$
with radial order $n=-1,\cdots,-8$ as a function of the period of the mode 
($|n|$ increases with the period). $B_{\l n} $:  
The modes with the same degree are represented by the same 
symbol: $\l = 1$ (red bullet), $\l = 2$ (blue square), and   $\l = 3$ (green diamond). 
The dipole modes $g_{-2}$ and $g_{-3}$
 have larger values due to 
a  {\it resonance} effect.}
\label{fig:3}
\end{figure}

In Figure~\ref{fig:2}, we show that $\Psi_{\l n}$ have non-null values only in the
region where the $^8B$ neutrino flux is produced. The form of the $\Psi_{\l n}$ 
depends  on the location of the maximum of $ \left({\delta T}/{T}\right)_{\l n}$
within the $^8B$ neutrino production region.  The largest amplitudes of the {\it gravito-neutrino eigenfunction} 
correspond to dipole $g$-modes of low $n$.
Eigenmodes $\left({\delta T}/{T}\right)_{\l n}$ with large $\l$ or $|n|$ 
have a very small impact on the production of solar neutrinos and
${\cal B}$ decreases rapidly with $\l$ (see Table~\ref{tab:1} and Figure~\ref{fig:3}).  
 
\section{Estimated effects and Discussion}

The  variation of the $^8B$ neutrino flux due to a given $g$-mode  
depends on the excitation source ${\cal A}_{\l n}$ and of the form factor
${\cal B}_{\l n}$ of each eigenfunction (Equation~\ref{eq:B_ln}).
If we consider a value of ${\cal A}_{\l n}$ ($\equiv |\delta T/{T}|_{\l n,\rm max}$) 
of $10^{-3}$, therefore, from Equation (\ref{eq:delta_phi_t})
for each eigenmode the maximum value of $\Delta \phi_{ln}(t)$ 
is such that  $|\Delta \phi/\phi|_{\l n,\rm max} =25.5\;|{\cal A}|_{\l n} \; |{\cal B}|_{\l n}$
(see Table~\ref{tab:1}). We found that the unique 
maximum of $\left(\delta T/{T}\right)_{\l n}$ for the dipole mode $g_{-2}$ occurs near the maximum 
of the $^8B$ neutrino source (see Figure~\ref{fig:1}). From the proximity between these two maxima, a resonance occurs
and $|\Delta \phi/\phi|_{\l n,\rm max}$  varies by 17 \%, which produces an amplification of 170 in comparison with the fluctuation of temperature for this specific mode. At least six other modes present amplitudes between 4 \% and 12\% for the same fluctuation of temperature.
 
\begin{table}[t]
\centering
\caption{$^8B$ Neutrino Flux Variations \\ Caused by $g$-Modes for a Fluctuation of $|{\Delta T \over T}|$ = 0.001.}
\begin{tabular}{clllllc}
\hline
\hline
 Degree &  Order &  Freq.& ($\mu Hz$)     & Period    &   $B_{\l n} $  & $|\Delta \phi/\phi|_{\l n,\rm max}$   \\
  $\l$  &   $n$     &  SSM & SeSM &   (hr)   &  ($-$)  &  (\%)  \\
 \hline
$1 $ &-1      & $254.7 $&  262.7 &  $1.1$    &  $2.3$   & $5.8$ \\
$-$  &-2      & $185.5$  &  191.1& $1.5$    &  $6.7$    & $17.0$ \\
$-$  &-3      & $148.6 $  & 153.2& $1.9$    &  $4.8$   & $12.2$ \\
$-$  &-4      & $123.6 $  & 127.8& $2.2$  &  $2.5$     & $6.4$ \\
$-$  &-5      & $105.6 $  &  109.3& $2.6$   &  $1.1$   & $2.8$ \\
\hline
$2$  &-1      & $291.5 $  & 296.4& $0.95$    & $1.0$  & $2.6$ \\
$-$  &-2      & $250.3 $  &  256.1&$1.1$  &  $1.5$    & $3.8$ \\
$-$  &-3      & $216.8 $  &  222.0&$1.3$  &  $1.6$    & $4.0$ \\
$-$  &-4     & $188.6 $  & 194.1& $1.5$   &  $1.7$    & $4.3$ \\
$-$  &-5     & $165.3 $  &  151.3& $1.7$   &  $1.5$    & $3.8$ \\
\hline
$3$  &-1      & $333.6 $  &340.1&  $0.83$  &  $1.6$   & $4.1$ \\
$-$  &-2      & $288.8 $  &  296.6& $0.96$   & $1.9$   & $4.8$ \\
$-$  &-3      & $255.9 $  &  261.3 &$1.1$  &  $1.0$   & $2.6$ \\
$-$  &-4     & $233.6 $  &  238.3&$1.2$   &  $0.9$    & $2.3$ \\
$-$  &-5      & $211.2 $  &  217. 1&$1.3$   &  $1.5$   & $3.8$  \\
\hline
\hline
\end{tabular}
\label{tab:1}
\vspace{0.5cm}
\end{table}

Of course, the variability of the boron flux strongly depends on the central amplitude of the gravity modes. Up now, no boron flux variability has been shown, so one can deduce an upper  estimate of temperature fluctuation. The Sudbury Neutrino Observatory team \citep{2010ApJ...710..540A} has analyzed a time series of 
$^8B$ neutrino fluxes to infer if any kind of periodic oscillation could be found; they look for any period from hour to days. They conclude that no statistically significant periodic signal was detected and that the effect, if any, must be smaller than 10\%.
From Equation (\ref{eq:delta_phi_t}), it follows that 
$|\Delta\phi/\phi|_{\l n}=25.5\; |{\cal A}|_{\l n}\,|{\cal B}|_{\l n,\rm ssm }$,
as current neutrino observations set an upper limit $|\Delta\phi/\phi|_{\l n,\rm max} = 0.10$, 
we determine that the maximum temperature fluctuation deduced from the $g_{-2}$ dipole mode 
to be $|\delta T/T|_{\l n,\rm max} \le 5.8\times 10^{-4}$ as $|{\cal B}_{1,-2}|_{\rm ssm } = 6.7$
(see Table~\ref{tab:1}).

$G$-mode dipole velocity surface amplitudes have been estimated at the level of few millimetres per second~\citep{2009A&A...494..191B}, in reasonable agreement with the GOLF first manifestation of $g$-mode detection at the surface \citep{2004ApJ...604..455T}. However, it is not so easy to properly estimate the amplitude of $g$-modes 
at the base of the convection zone or in the core. It is note even proven that these modes are only excited by the convection at the base of the convective zone, as they can also be excited by the opacity bump of hydrogen or helium \citep{2004ApJ...613L.169C}.
Nevertheless, the mechanisms by which turbulent convection excites and affects the propagation 
of gravity waves in the  Sun has been extensively discussed in the 
literature~\citep{1996ApJ...458L..83K,1977ApJ...212..243G,1991ApJ...374..366G,1996A&A...312..610A,1975PASJ...27..401S}.
These turbulent motions at BCZ very likely excite gravity waves. Following~\cite{1993ApJ...409L..73B},  ${\cal A}_{\rm core}  \sim 10^{-7} V_{\rm core} $  
where the velocity of the mode $V_{\rm core}$ is in  ${\rm cm s^{-1}}$.
This calculation leads to  some temperature fluctuation ${\cal A}_{\rm core}$ of the order of 10$^{-5}$, but this estimate could be pessimistic as their calculations underestimate the gravity mode velocity at the surface.
\citet{1996A&A...312..610A} found in numerical simulations that
although the turbulent motions at the lower part of the convection zone
produce $g$-modes of relatively large amplitude beneath the BCZ,
due to the attenuation caused by the interaction of $g$-modes
with the structure in the convection zone,
the amplitudes  of large $\l$ modes at the solar surface are quite small; 
turbulent motions with a $V_{\rm bcz}\sim 50\;{\rm m s^{-1}}$ 
produce $g$-modes of moderate degree with an 
amplitude velocity at the Sun's surface of the order of $0.01$ -- $5\; {\rm mm s^{-1}}$.
They predict that $g$-modes of low degree should have surface amplitudes slightly larger and consequently observable, in agreement with \cite{Garcia2007} detections.
More recently, \citet{2012ApJ...757..128M}, based on magneto-hydrodynamics considerations estimate that  
the velocity of turbulent motions ($V_{\rm bcz}$) at the base of the convection  zone (BCZ)  
is of the order of  $10\;{\rm m s^{-1}}$ and confirm them by numerical simulations. The three-dimensional (3D) simulations will be useful to estimate the temperature fluctuations but they are not yet sufficiently reliable to produce some relevant numbers for this study. 

We orient attention toward $^8B$ neutrino fluxes for the new generation of detectors and show the great interest to  look in the  core to better determine the gravity mode properties. It represents a complementary approach to helioseismology. Such a search could be also extend to the effect of gravity waves of lower frequencies and will benefit from the development of 3D simulations of the entire Sun. These non-standing waves describe a spiral in the radiative zone and lose their energy before reaching the core and thus could also impact  the pp flux~\citep{2014A&A...565A..42A}. Improving the statistics  of pp flux detectors  is also of high interest for precisely determining the solar energy production.
The different approaches discussed here suggest that the core amplitude temperature perturbations 
must be found below 10$^{-3}$ but above 10$^{-5}$; therefore they will be detectable with the new generation of neutrino detectors like SNO+ and Hyperkamiokande.

\section{Summary and Conclusion}
 
In this Letter a new strategy for the search of $g$-modes inside the solar interior is proposed.
Due to the high sensitivity of neutrino nuclear reactions 
to the temperature, the $^8B$ neutrino flux time series is an ideal tool to probe the existence of $g$-modes inside the Sun.
This $^8B$ neutrino flux should vibrate with the same frequencies as the $g$-modes 
which produce  local temperature changes. Thus, these time series should also contain information on the excitation and dumping of gravity waves in the solar core.   

We found that only a few $g$-mode amplitudes are largely amplified in the $^8B$ neutrino flux. 
Actually,  $g$-modes with low degrees ($\l\le 3$)  have the strongest impact on the $^8B$ neutrino flux. 
For dipole $g$-modes ($\l$ with $n=-2$),  the temperature perturbation is amplified by a factor of 170,  practically a factor 10 greater than the normal sensitivity to the temperature of this neutrino flux. 
In fact,  the maxima 
of the temperature eigenfunctions of the dipole modes  with $n = -2$ and $n = -3$ are located near the maximum  
of the source of the $^8B$ neutrino (see Figure~\ref{fig:1}).   The present study orients  the experimental search of $g$-modes 
on the period interval between 0.5 and 4hr, where the $g$-modes induce the strongest time variation on the $^8B$ neutrino flux. The research will be particularly well adapted for the coming generation of neutrino detectors like SNO+ and Hyperkamiokande for which the statistics will be multiplied by 25. This progress on detection will be extremely useful to better constrain the neutrino properties, solar central temperature, some ingredients of the SSM, or even extra processes. As mentioned in this paper, this  will also help  to look for time variability of the boron neutrino flux, so that if this detection is positive, it will be determinant for exploring the very deep dynamics of the Sun.

Considering the analysis by~\cite{2010ApJ...710..540A} of SNO time series of 
$^8B$ neutrino flux, which concluded that no statistically significant periodic signal was detected, 
and fixing an upper limit 
of 10 \% of potential variability of $^8B$ neutrino flux, we put an upper limit of the fluctuations of the central solar temperature of $5\times 10^{-4}$.

The possible discovery of $g$-modes in the $^8B$ neutrino flux could revolutionize the field of solar physics. It will provide 
a new tool to probe the Sun's core with unprecedented details. 
At a minimum, such experimental analysis  will put  important constraints 
on the maximum amplitude of $g$-modes in the solar core. In addition, such study will put new constraints on the low-frequency part of the inertial gravity waves (periods smaller than four days) which are not detectable with seismology. 
This new quantified approach will be extended to other neutrino fluxes.

\begin{acknowledgments}
We thank the anonymous referee for constructive remarks. 
This work was supported by grants from "Funda\c c\~ao para a Ci\^encia e Tecnologia" 
and "Funda\c c\~ao Calouste Gulbenkian".
\end{acknowledgments}
%

%
\end{document}